\documentstyle[prl,aps,twocolumn,psfig]{revtex}
\begin{document}
\flushbottom
\draft
\title{Optimal universal two-particle
entanglement processes in arbitrary dimensional Hilbert spaces}
\author{G.~Alber$^1$, A.~Delgado$^1$, and I.~Jex$^{1,2}$} 
\address{
$^1$Abteilung f\"ur Quantenphysik, Universit\"at Ulm,
D--89069 Ulm, Germany\\
$^2$ Department of Physics, FJFI \v CVUT,
B\v rehov\'a 7, 115 19 Praha 1 - Star\'e M\v{e}sto, Czech Republic\\
(submitted to Phys. Rev. A.)\\
}
\date{\today}
\maketitle
\begin{abstract}
Universal two-particle entanglement processes are analyzed in arbitrary
dimensional Hilbert spaces. On the basis of this analysis the class of
possible optimal universal entanglement processes is determined
whose resulting output states do not contain any separable states.
It is shown that these processes form a one-parameter family.
For all Hilbert space dimensions larger than two the resulting
optimally entangled output states are mixtures of anti-symmetric
states which are freely entangled and which also
preserve information about input states. Within this one-parameter
family there is only one process by which all information about
any input state is destroyed completely.
\end{abstract}

\section{Introduction}

One of the main driving forces in the rapidly developing field of quantum information
processing is the question whether basic quantum phenomena such as interference
and entanglement can be exploited for practical purposes. In this context
it has been realized that
the linear character of quantum theory may impose severe restrictions on the
performance of elementary tasks of quantum information processing.
As a consequence it is impossible to copy (or clone) an arbitrary quantum state
perfectly \cite{Wooters}.
In view of the significance of entangled states for many aspects of
quantum information
processing the natural question arises whether similar restrictions also hold for
quantum mechanical
entanglement processes. Of particular interest are entanglement processes which
entangle two quantum systems
in an optimal way
in the sense that the corresponding two-particle quantum state 
does not contain any separable components.
Though many quantum mechanical processes are capable of entangling some input
states of a quantum system
with a known reference state of a second quantum system, it is not easy to achieve this goal for
all possible input states. This basic difficulty
can be realized already in the simple example of a
quantum mechanical controlled-not (CNOT) operation, i.e. 
\begin{eqnarray}
{\rm CNOT}: |\pm\rangle \otimes (|+\rangle + |-\rangle) &\to& |\mp\rangle \otimes |+\rangle +
|\pm\rangle \otimes |-\rangle. \nonumber
\end{eqnarray}
This CNOT operation entangles the orthogonal input states $|\pm\rangle$ of the first qubit
with the second (control) qubit prepared in the reference state 
$(|+\rangle + |-\rangle)$. Obviously the two Bell states resulting from these input states
are optimally entangled. However, due to its linearity
this quantum process is incapable of entangling the first qubit
with the second one for all possible input states. The input state
$(|+\rangle + |-\rangle)$, for example, results in the factorizable output state
$(|+\rangle + |-\rangle)\otimes (|+\rangle + |-\rangle)$.
In view of this difficulty it is of particular
interest to investigate universal entanglement processes
which are able to entangle all possible input states
of a quantum system with a second one in an optimal way. 

Universal quantum processes act on all possible (typically pure) input states of a 
quantum system in a `similar' way. Consequently, these processes do not specify
a preferred direction
in Hilbert space and thus reflect its `natural' symmetry.
Therefore, the restrictions imposed on these processes by the linear character
of quantum theory are not only of practical interest but they also hint
at fundamental limits of quantum theory.
So far many properties of universal quantum processes
have been analyzed for qubits \cite{clone,Buzek}.
For qubits one can show that there is only one universal,
optimal entanglement process. Independent of the input states this process
produces the anti-symmetric Bell state as an optimally
entangled output state \cite{Alber}.
However, for many applications in quantum information processing
universal quantum processes are needed
which do not only entangle different quantum systems in an optimal way
but which also preserve information about the original
input state and redistribute this information into
an entangled two-particle state.  
Motivated by this need recently 
Bu\v zek and Hillery \cite{Buzek} have analyzed various quantum processes
which entangle two qubits and 
which also preserve information about the initial input state. 
Though for universal quantum processes 
both requirements are incompatible in the case of qubits
universal optimal cloning processes manage to optimize both 
tasks simultaneously. However, the resulting output states
are not optimally entangled as they
always contain a separable two-qubit state.
From these investigations on qubit systems
one may be tempted to presume that a similar incompatibility
between optimal entanglement and preservation of information about
input states also holds in higher dimensional Hilbert spaces.

In this paper it is shown that, contrary to this tempting presumption,
in Hilbert spaces of dimensions higher than two
optimal universal entanglement processes are possible which preserve information about the
initial input state and which also lead to optimally entangled output states.
For this purpose a convenient theoretical framework is developed which is capable of describing
all possible universal quantum processes involving two quantum systems.
For the sake of simplicity
we restrict our discussion to the important special case that the dimensions of
the Hilbert spaces of both quantum systems are equal.
First of all,
the general class  of all possible universal quantum operations is determined
which is
compatible with the linear character of quantum mechanics. Secondly,
the particular subclass is determined which
produces
optimally entangled two-particle output states in the sense that these output
states do not contain any separable components.
It turns out that for Hilbert spaces with dimensions larger than two
these optimal universal entanglement processes form a one-parameter family.
Within this one-parameter family
there is only one particular process 
producing
optimally entangled output states which are independent of the input states. This
particular optimal universal entanglement process
has already been discussed previously \cite{Alber}. All other processes within this
one-parameter family lead to 
output states which also contain information about the input
state. One of these processes 
preserves this information about the original input
state in an optimal way.
It turns out that
all the resulting optimally entangled output states are anti-symmetric with respect
to particle exchange.

This paper is organized as follows: In Sec. II the basic symmetry (or covariance) property
of universal quantum processes is discussed by starting from 
a simple example.
subsequently a general formalism is developed for
describing all possible universal quantum processes in arbitrary dimensional Hilbert spaces.
The consequences of covariance and of the linear character of universal quantum processes
are implemented.
In Sec. III all possible optimal universal entanglement processes are determined and
basic properties of the resulting output states are investigated.

\section{Universal quantum processes}

In this section the basic symmetry property of universal quantum processes 
is exemplified by considering a simple example involving a process with two qubits. 
Based on this symmetry property and on the requirement that any quantum process
has to be linear with respect
to all possible input states the general structure of universal quantum processes is
discussed for the case of two arbitrary dimensional quantum systems.
Optimal universal quantum cloning processes and optimal universal entanglement processes
are special cases thereof.

\subsection{Symmetry of universal quantum processes - an example}
Let us consider the following quantum process as an introductory example:
\\
Initially we prepare two distinguishable spin-1/2 quantum systems
(qubits) in the state
$$\rho_1({\bf m})\equiv\rho_{in}({\bf m})\otimes \frac{1}{2}{\bf 1}.$$
The pure input state $\rho_{in}({\bf m}) = |{\bf m}\rangle \langle {\bf m}|$
of the first quantum system can be described by its Bloch vector ${\bf m}$.
This Bloch vector can take an arbitrary position on the Poincare sphere.
The second quantum system is in a completely unpolarized mixed reference state
which is assumed to be fixed once and for all.
Selecting an arbitrary pure input state $\rho_{in}({\bf m})$ we 
transfer the initial state $\rho_1({\bf m})$ into the output state
\begin{equation}
\rho_1({\bf m}) \to \rho_2({\bf m}) = \frac{{\bf P}_J \rho_1({\bf m}) {\bf P}_J}
{{\rm Tr}{\bf P}_J \rho_1({\bf m}) {\bf P}_J}.
\label{example}
\end{equation}
Thereby the projection operator
${\bf P}_J = \sum_M|J M\rangle \langle J M|$
projects onto two-particle states with well defined
total angular momentum $J$. 
This total angular momentum can assume the possible values
$J=1$ or $J=0$.
In a probabilistic way
the transformation of Eq.(\ref{example}) can be achieved
by a measurement process. 
However, one may also think of realizing this transformation 
with a probability of unity in a unitary
and reversible way, for example. 
Choosing the direction of polarization of the input state as the
quantization axis the result of this quantum process is 
given by
\begin{eqnarray}
\rho_2({\bf m}) &=& p_1 |J=1 M=1\rangle \langle J=1 M=1| +\nonumber\\
&& (1 - p_1)
|J=1 M=0\rangle\langle J=1 M=0|
\label{clone}
\end{eqnarray}
with $p_1 = 2/3$
or by
\begin{equation}
\rho_2({\bf m}) = |J=0 M=0\rangle \langle J=0 M=0|
\label{entangle1}
\end{equation}
depending on whether $J=1$ or $J=0$.
This quantum process 
is universal in the sense that all input states are treated
in a `similar' way. The only direction the output state depends on is
the one of the input state.
Thus this quantum process 
is symmetric with respect to unitary transformations $U$ which transform
an arbitrary pure one-particle input state, say $|{\bf m}_0\rangle$, into some other
pure one-particle input state, say $|{\bf m}\rangle \equiv U|{\bf m}_0\rangle$.
This unitary symmetry or covariance of such a universal quantum process
is characterized by the relation 
\begin{equation}
\rho_2({\bf m}) = 
U({\bf m})\otimes U({\bf m})
\rho_2({\bf m}_0)
U^{\dagger}({\bf m})\otimes U^{\dagger}({\bf m})
\label{covariance}
\end{equation}
(compare with Fig. \ref{Fig1}). Thus
the possible output states of a universal quantum process constitute a two-particle
representation
of the group of unitary one-particle transformations.

Universal quantum processes in which the step of Eq.(\ref{example})
can be implemented 
with a probability of unity
have been
investigated in the context of copying (cloning) quantum states.
In particular, it has been demonstrated that optimal quantum cloning can be achieved always
by a universal quantum process. Furthermore, in the case of two qubits
the maximum probability with which an optimal
universal quantum cloning process is successful is given by $2/3$ \cite{clone}. This latter probability
is identical with the probability $p_1$ appearing in Eq.(\ref{clone}).
Thus, provided the process of Eq.(\ref{example})
is implemented for $J=1$ and with a probability of unity this process
copies an arbitrary input state in an optimal way.
However, if we project onto states with $J=0$, 
we end up in the anti-symmetric Bell state formed by both qubits. Furthermore, this
output state is independent of the input state which we choose.
As a Bell state is maximally entangled this latter type of process is an example of
a universal optimal entanglement process.

Copying quantum states and preparing entangled quantum states are elementary tasks
of quantum information processing. Thereby
universal quantum processes fulfilling Eq.(\ref{covariance})
which exhibit the same symmetry as the set of all
possible pure one-particle input states are of special interest. 
Though much is already known about universal quantum cloning
processes almost nothing is known about universal quantum processes
which yield optimally entangled quantum states, in particular in arbitrary dimensional
Hilbert spaces.
The main questions which will be addressed in the following are:
Is it possible at all to produce optimally entangled quantum states
by universal quantum processes? Which limitations are imposed on the structure
of these states by the universality and linearity of such a quantum process?
How do the properties of resulting optimally entangled states depend on the
dimensionality of the Hilbert spaces involved?

\subsection{General structure of universal quantum processes involving two quantum
systems}
Let us
consider the most general universal quantum process which is capable of
entangling two quantum systems, i.e.
\begin{equation}
{\cal P}: \rho_{in}({\bf m})\otimes \rho_{ref} \to \rho_{out}({\bf m}).
\label{universal}
\end{equation}
In our previous example the fixed reference state $\rho_{ref}$
was maximally mixed. In the present
case we leave its form unspecified.
The density operator
of the pure input state
is denoted
$\rho_{in}({\bf m})$.
For the sake of simplicity let
us assume that the dimensions of the Hilbert spaces for both quantum 
systems are equal and of magnitude $D$.
In order to classify all possible universal quantum processes of the
form of Eq.(\ref{universal}) we have to determine the most general form
of the input and output states.

The density operator of an
arbitrary input state of a $D$ dimensional quantum system
can always be represented in terms of the
generators ${\bf A}_{ij}$ $(i,j=1,...,D)$ of the group $SU_D$, i.e. 
\begin{equation}
\rho_{in}({\bf m})=\frac{1}{D}({\bf 1} + m_{ij}{\bf A}_{ij}).
\label{input}
\end{equation}
(We use the Einstein summation convention in which one has to sum over all indices which
appear in an expression twice.)
A representation of these generators is given by
the $D\times D$ matrices
\begin{equation}
({\bf A}_{ij})^{(kl)} = \delta_{ik}\delta_{jl} -
\frac{1}{D}\delta_{ij}\delta_{kl}. 
\label{repr}
\end{equation}
These matrices are not hermitian but they fulfill the relation
${\bf A}_{ij}^{\dagger} = {\bf A}_{ji}$. Due to the constraint
$\sum_{i=1}^{D}{\bf A}_{ii} = 0$
only $(D^2-1)$ of them are linearly independent.
For D=2 these matrices reduce to the well known spherical components of the
Pauli spin matrices, i.e. $2{\bf A}_{11}=\sigma_z$,
$2{\bf A}_{12}=\sigma_x + i\sigma_y$ and 
$2{\bf A}_{21}=\sigma_x - i\sigma_y$.
Furthermore, $\rho_{in}({\bf m})=\rho_{in}({\bf m})^{\dagger}$ 
implies the relations $[m_{ij}]^*=m_{ji}$
so that Eq.(\ref{input}) involves
$(D^2-1)$ real-valued and linearly
independent parameters $m_{ij}$ which form the components of a
generalized Bloch vector. This Bloch vector is an observable
of the quantum system which defines its state uniquely.
Eq.(\ref{input}) represents a pure state provided 
${\rm Tr}(\rho_{in}({\bf m})^2) = 1$ which is equivalent to the constraint
$\sum_{i,j}|m_{ij}|^2 - (1/D) (\sum_{i}m_{ii})^2 = D(D-1)$.

In terms of the generators of Eq.(\ref{repr}) the most 
general two-particle output state
is represented by a density operator of the form
\begin{eqnarray}
\rho_{out}({\bf m}) &=& \frac{1}{D^2}{\bf 1}\otimes {\bf 1}
+ \alpha_{ij}^{(1)}({\bf m}) {\bf A}_{ij}\otimes {\bf 1} +
\nonumber
\\
&& 
\alpha_{ij}^{(2)}({\bf m})  {\bf 1} \otimes {\bf A}_{ij} +
K_{ijkl}({\bf m}){\bf A}_{ij}\otimes {\bf A}_{kl}.
\label{two-general}
\end{eqnarray}
In order to implement the covariance condition (\ref{covariance})
it is useful to separate the last term of Eq.(\ref{two-general})
into terms which are invariant and into terms which transform as the generators
${\bf A}_{ij}$
under arbitrary unitary transformations of the form $U\otimes U$.
For this purpose it is useful to start from the commutation relations
of $SU_D$, namely
\begin{equation}
[{\bf A}_{ij},{\bf A}_{mn}] = {\bf A}_{ab}
(\delta_{jm}\delta_{ai}\delta_{bn} -
\delta_{in}\delta_{am}\delta_{bj}). 
\label{commutation}
\end{equation}
These relations imply
that the tensor products
${\bf A}_{ji}\otimes{\bf A}_{sj}$ transform under arbitrary
transformations of the form $U\otimes U$ in the same way as ${\bf A}_{si}$ 
transforms under transformation of the form $U$. Furthermore,
the tensor product ${\bf A}_{ij}\otimes{\bf A}_{ji}$ is an invariant
under arbitrary unitary transformations  of the form $U\otimes U$.
However, note that the combination 
${\bf A}_{ij}\otimes{\bf A}_{sj}$, for example, 
does not transform analogous to ${\bf A}_{si}$.
Using these elementary transformation properties, the covariance condition
(\ref{covariance}), and the fact that any quantum operation has to
be linear with respect to its input states
we obtain the 
most general form for the 
density operator of the
two-particle output state, namely 
\begin{eqnarray}
\rho_{out}({\bf m}) &=& \frac{1}{D^2}{\bf 1}\otimes {\bf 1}
+ \alpha_{ij}^{(1)}({\bf m}) {\bf A}_{ij}\otimes {\bf 1} +
\nonumber\\
&&
+ \alpha_{ij}^{(2)}({\bf m})  {\bf 1} \otimes {\bf A}_{ij} +
C {\bf A}_{ij}\otimes {\bf A}_{ji} + 
\nonumber
\\
&& 
\beta_{il}({\bf m}) {\bf A}_{ij}\otimes {\bf A}_{jl} + 
\beta_{il}({\bf m})^* {\bf A}_{ji}\otimes {\bf A}_{lj} 
\label{ouput}
\end{eqnarray}
with
\begin{eqnarray} 
\alpha_{ij}^{(1,2)} &=& \alpha^{(1,2)} m_{ij},\nonumber\\
\beta_{ij}&=& \beta m_{ij}
\label{output1}
\end{eqnarray} 
and with $C$ being independent of ${\bf m}$.

So far the output state of Eq.(\ref{ouput})
represents the most general hermitian operator which depends
linearly
on the input state $\rho_{in}({\bf m})$ and which fulfills the covariance condition
(\ref{covariance}).
Accordingly,
a particular universal quantum process is characterized
by the set of
real-valued parameters
$C$, $\alpha^{(1)}$, $\alpha^{(2)}$ and by the complex valued parameter
$\beta$.
We still have to solve the more difficult task to restrict the 
range of these parameters 
in such a way that $\rho_{out}({\bf m})$ of Eq.(\ref{ouput})
represents a non-negative operator. 
In order to determine this fundamental range of these parameters
we have to investigate the possible eigenvalues of the density operator
$\rho_{out}({\bf m})$ of Eq.(\ref{ouput}). 
Due to the covariance condition (\ref{covariance}) 
we may restrict this investigation to a particular pure input state 
with $m_{ij} = D\delta_{i1}\delta_{j1}$, for example. 
Using the matrix representations of Eq.(\ref{repr}) it turns out that
the corresponding
output state can be represented by a direct sum 
of density operators according to 
\begin{eqnarray}
\rho_{out}({\bf m}\equiv D{\bf e}_{11})
&=&
\sum_{i=1}^{4}\oplus p_i \rho_{i} 
\label{rhoout}
\end{eqnarray}
with the partial density operators
\begin{eqnarray}
\rho_{1} &=&|11\rangle \langle 11|,\nonumber\\
\rho_{2} &=&
\sum_{j=2}^D \{|1j\rangle \langle 1j|
(\frac{1}{2(D-1)} +
\frac{(\alpha^{(1)} - \alpha^{(2)})m_{11}}{2p_2}) +\nonumber\\
&&
|j1\rangle \langle j1|
(\frac{1}{2(D-1)} +
\frac{(\alpha^{(2)} - \alpha^{(1)})m_{11}}{2p_2})
+\nonumber\\
&&
|1j\rangle \langle j1|\frac{C + \beta m_{11}}{p_2} +
|j1\rangle \langle 1j|\frac{C + \beta^* m_{11}}{p_2} \}, \nonumber\\
\rho_{3} &=&\frac{1}{(D-1)}
\sum_{j=2}^D |jj\rangle \langle jj|,\nonumber\\
\rho_{4} &=&
\sum_{2=i<j}^D \{|ij\rangle \langle ij|\frac{1}{(D-1)(D-2)}
+\nonumber\\
&&|ji\rangle \langle ji|\frac{1}{(D-1)(D-2)}
+\nonumber\\
&&
|ij\rangle \langle ji|\frac{C}{p_4}  + |ji\rangle \langle ij|\frac{C}{p_4}\}.
\label{explicit}
\end{eqnarray}
The corresponding partial probabilities entering Eq.(\ref{rhoout}) are given by
\begin{eqnarray}
p_1 &=&
\frac{1}{D^2} + (\alpha^{(1)} + \alpha^{(2)})m_{11}(1-\frac{1}{D}) +
C(1 - \frac{1}{D}) + \nonumber\\
&&
(\beta + \beta^*) m_{11} (1 - \frac{1}{D})^2,\nonumber\\
p_2 &=&(D-1)\{
\frac{2}{D^2} + (\alpha^{(1)}+\alpha^{(2)})m_{11}(1-\frac{2}{D}) 
-\nonumber\\
&& \frac{2C}{D} - 2(\beta + \beta^*)m_{11}(1 - \frac{1}{D})\frac{1}{D}\},
\nonumber\\
p_3 &=&(D-1)\{
\frac{1}{D^2} 
- \frac{\alpha^{(1)}m_{11}}{D}
- \frac{\alpha^{(2)}m_{11}}{D} + \nonumber\\
&&C(1-\frac{1}{D}) + (\beta + \beta^*)m_{11}\frac{1}{D^2}
\},\nonumber\\
p_4 &=& (D-1)(D-2)\{
\frac{1}{D^2} 
- \frac{\alpha^{(1)}m_{11}}{D}
- \frac{\alpha^{(2)}m_{11}}{D} - \nonumber\\
&&\frac{C}{D} + (\beta + \beta^*)m_{11}\frac{1}{D^2}
\}.
\label{prob}
\end{eqnarray}
The normalization of the density operator, i.e.
${\rm Tr}(\rho_{out}({\bf m}))=1$, implies 
\begin{equation}
p_1 + p_2 + p_3 + p_4 = 1.
\label{conservation}
\end{equation}
From Eqs.(\ref{explicit}) and (\ref{prob}) one obtains the eigenvalues
of $\rho_{out}({\bf m}=D{\bf e}_{11})$, namely 
\begin{eqnarray} 
\lambda_{1}&=&p_1,\nonumber\\
\lambda_{2\pm} &=&\frac{p_2}{2(D-1)} \pm
\sqrt{(\frac{(\alpha^{(1)}-\alpha^{(2)})m_{11}}{2})^2
+ \mid C + m_{11}\beta \mid^2}, \nonumber\\
\lambda_{3}&=&\frac{p_3}{(D-1)},\nonumber\\
\lambda_{4\pm}&=&\frac{p_4}{(D-1)(D-2)} \pm \mid C\mid.
\label{eigen}
\end{eqnarray} 
Therefore the density operator of Eq.(\ref{rhoout}) is non-negative
only if all probabilities $p_i$ and all eigenvalues
$\lambda_i$ of Eqs.(\ref{prob}) and (\ref{eigen}) are non-negative and
fulfill Eq.(\ref{conservation}).
For 
$\alpha^{(1)}=\alpha^{(2)}$ and $\beta = \beta^*$, for example,
these conditions on $(p_2,p_3,p_4)$ form
a tetrahedron (compare with Fig.\ref{Fig2}).
Each point in this convex set defines
a different universal quantum process whose
possible output states can be obtained from Eq.(\ref{rhoout}) with the help of the
covariance condition (\ref{covariance}).
The
universal quantum cloning process, for example, is represented by point $B$ in this figure
and it is characterized by this particular universal process which maximized $p_1$.
Note that it is immediately obvious from Fig.\ref{Fig2} that perfect quantum cloning is
impossible with a universal quantum process as
$p_1 = 1 - p_2 - p_3 - p_4 \leq 2/(D+1) <1$.

Finally, it should be mentioned that for dimensions
$D\geq 3$ one may choose
the probabilities $(p_1,p_3,p_4)$ or $(p_2,p_3,p_4)$, for example,
as independent coordinates 
instead of
the three independent real-valued parameters 
$((\alpha^{(1)}+\alpha^{(2)}),C,(\beta+\beta^*))$.
Inverting Eqs.(\ref{prob}) and using Eq.(\ref{conservation})
one obtains the relation between these different
coordinate systems, namely
\begin{eqnarray}
\beta+\beta^* &=& -\frac{1}{D(D-1)}+\frac{p_4}{(D-1)(D-2)}+
\frac{p_1}{(D-1)},\nonumber\\
\alpha^{(1)}+\alpha^{(2)} &=& \frac{(D-2)}{2D^2(D-1)}
-\frac{p_4}{2D(D-1)}+\frac{p_1}{2D^2(D-1)} -\nonumber\\
&&\frac{p_3}{2D(D-1)},\nonumber\\
C&=&\frac{p_3}{(D-1)}-\frac{p_4}{(D-1)(D-2)}.
\label{invert}
\end{eqnarray}
In order to identify a particular universal quantum process uniquely in addition
to these three probabilities one also
has to specify the remaining two independent parameters, namely $(\alpha^{(1)} - \alpha^{(2)})$
and $(\beta - \beta^*)$.

\section{Optimal universal entanglement processes}

Starting from the notion of optimal entanglement as defined by
Lewenstein and Sanpera \cite{Lewenstein}
it is shown that there is a unique one-parameter family of optimal universal
entanglement processes for Hilbert spaces with dimensions larger than two.
All these processes produce output states which are anti-symmetric with respect
to particle exchange.
Characteristic properties of the resulting
output states are investigated. It is demonstrated that among this one-parameter
family of optimal entanglement processes there is only one process in which
all information about input states is lost. All other processes preserve this information
at least partly. Within this one-parameter family 
there is also one particular process which preserves this information
optimally.

\subsection{Characterization of optimal universal entanglement processes}

Is it possible to entangle two quantum systems in an optimal way
by a universal quantum process?
Before addressing this question one has to clarify the meaning of 
optimal entanglement. As discussed by Lewenstein and Sanpera \cite{Lewenstein}
one can decompose any quantum state $\rho$ of a composite
system into a separable
part, say $\rho_{sep}$, and an inseparable contribution $\rho_{insep}$, i.e.
$\rho = \lambda \rho_{sep} + (1 - \lambda)\rho_{insep}$ with $0\leq \lambda \leq 1$.
Thereby a separable state is a convex sum of product states of the form $\rho_A\otimes
\rho_B$ where $\rho_A$ and $\rho_B$ refer to quantum
systems $A$ and $B$ respectively.
Though this decomposition itself is not unique the maximum value of
$\lambda$ is.
Thus a quantum state may be called
optimally entangled if the maximum possible value
of $\lambda$ equals zero in any such decomposition.

In order to determine the parameters for the universal quantum processes which
produce optimally entangled states 
let us start from the output state
$\rho_{out}({\bf m}=D{\bf e}_{11})$ of Eq.(\ref{rhoout}).
A necessary condition for this state
being optimally entangled is
that there are no admixtures of separable
states of the form $|jj\rangle \langle jj|$ for any
$j=1,...,D$.
Thus, necessarily a universal quantum process producing
optimally entangled states 
has to be characterized by the parameters
\begin{eqnarray}
p_1 &=& 0,\nonumber\\
p_3 &=& 0. 
\label{entangle}
\end{eqnarray}
It will be demonstrated by the subsequent arguments that
this choice of parameters 
is also sufficient for the generation of optimally entangled
output states.
For this purpose it has to be proven that
for any separable two-particle state $|\psi\rangle = |\varphi\rangle \otimes |\chi\rangle$
and any positive value of $\lambda > 0$
the state
\begin{equation}
\rho'=\rho_{out}({\bf m}=D{\bf e}_{11})^{(opt)} - \lambda |\psi\rangle \langle \psi |
\label{test}
\end{equation}
is negative definite.
Thereby the state $\rho_{out}({\bf m}=D{\bf e}_{11})^{(opt)}$ fulfills
conditions (\ref{entangle}).
Due to the covariance condition (\ref{covariance})
this non-negativity then implies that also any arbitrary output state
$\rho_{out}({\bf m})^{(opt)}$ 
cannot contain any separable components. Thus, it is
optimally entangled.

For the proof of this statement we start from conditions
(\ref{entangle})
and Eqs.(\ref{prob}) and (\ref{eigen}).
According to Eqs.(\ref{eigen}) and (\ref{invert})
the condition $p_3 = 0$ implies $\lambda_{4-} = 0$.
Furthermore,
from the non-negativity of $\lambda_{2-}$ of Eq.(\ref{eigen}) and
from 
Eqs. (\ref{invert}) and (\ref{entangle})
we obtain the relations
\begin{eqnarray}  
\alpha^{(1)}&=&\alpha^{(2)},\nonumber\\
\beta &=& \beta^*,\nonumber\\
\rho_{2}^{(opt)}&=&\frac{1}{2(D-1)}
\sum_{j=2}^D \{|1j\rangle \langle 1j|
+
|j1\rangle \langle j1|
-\nonumber\\
&&
|1j\rangle \langle j1| -
|j1\rangle \langle 1j|\},\nonumber\\
\rho_{4}^{(opt)}&=&\frac{1}{(D-1)(D-2)}
\sum_{2=i<j}^D \{|ij\rangle \langle ij|
+|ji\rangle \langle ji|
-\nonumber\\
&&
|ij\rangle \langle ji|  - |ji\rangle \langle ij|\}.
\label{entangledstate}
\end{eqnarray}  
Thus, the parameters of Eqs.(\ref{entangle})
imply that the resulting output state
\begin{eqnarray}
\rho_{out}^{(opt)}({\bf m}=D{\bf e}_{11}) &=&(1-p_4)\rho_2^{(opt)}
\oplus p_4 \rho_4^{(opt)}
\label{optimal}
\end{eqnarray}
is a convex sum of two-particle quantum states which are anti-symmetric
with respect to permutations of both quantum systems.
Let us consider now the state $\rho'$ of Eq.(\ref{test}).
For an arbitrary state
$|\psi\rangle = |\varphi\rangle \otimes |\chi\rangle$
we can always choose a unitary transformation $U$
in such a way that
$\langle 1|U|\varphi\rangle$ and $\langle 1|U|\chi\rangle$ are both non-zero.
This unitary transformation may be interpreted passively as a change of basis
in the one-particle Hilbert spaces.
Applying the same unitary transformation to
state $\rho_{out}^{(opt)}({\bf m}=D{\bf e}_{11})$
produces a convex sum of anti-symmetric two-particle
states so that
$\langle 1 1|U\otimes U\rho_{out}^{(opt)} U^{\dagger}\otimes U^{\dagger}|1 1\rangle = 0$.
Thus, 
assuming the existence of a state
$|\psi\rangle = |\varphi\rangle \otimes |\chi\rangle$
and a probability $\lambda >0$
implies that for this particular unitary transformation
$U$
the diagonal density matrix element
$\langle 1 1|U\otimes U\rho' U^{\dagger}\otimes U^{\dagger}|1 1\rangle = 0 - \lambda
\langle 1|U|\varphi\rangle \langle 1|U|\chi\rangle$ is negative. Therefore $\rho'$ is
negative definite for any choice of the states $|\varphi\rangle$ and $|\chi\rangle$ and
for any $\lambda >0$.
Therefore a non-zero value of $\lambda$ is not possible in Eq.(\ref{test}).
So we conclude that the two-particle state
of Eq.(\ref{optimal}) is optimally entangled. By covariance the same property applies
to all possible output states.
This completes our proof.

\subsection{Basic properties of optimally entangled output states}

Thus, the 
parameters
\begin{eqnarray}
0\leq& p_4&\leq 1,\nonumber\\
p_1&=&0,\nonumber\\
p_3&=&0,\nonumber\\
\alpha^{(1)}&=&\alpha^{(2)},\nonumber\\
\beta&=&\beta^*
\end{eqnarray}
characterize all possible universal quantum processes
which produce optimally entangled two-particle output states.
The resulting output states 
are statistical mixtures of anti-symmetric states.
Explicitly they are given by Eqs.(\ref{entangledstate}), (\ref{optimal}) 
and by applying the covariance condition (\ref{covariance}).
In addition they exhibit other noteworthy
properties which will be discussed in the following.

The partial transpose of the
output state $\rho_{out}^{(opt)}({\bf m}= D{\bf e}_{11})$ 
of Eq.(\ref{optimal})
has always a negative eigenvalue of magnitude
\begin{eqnarray}
\Lambda &=& 
 -\frac{p_4}{2(D-1)}-
\{\frac{p_4}{(D-1)}^2 +\\
&& 4(D-1)[\frac{1}{2} - p_3\frac{D-2}{2(D-1)} -
\frac{p_4}{2} - p_2 \frac{D}{2(D-1)}]^2\}^{1/2}.\nonumber
\end{eqnarray}
Therefore, by covariance all optimally entangled states
which are produced by these universal entanglement processes are freely entangled
\cite{Horodecki}.

Due to covariance
all output states resulting from the same universal optimal entanglement process
have the same von Neumann entropy of magnitude
\begin{eqnarray}
S(p_4) &=& p_4 {\rm ln}\frac{(D-1)(D-2)}{2p_4} + (1 - p_4){\rm ln}\frac{(D-1)}{1 - p_4}.
\end{eqnarray}
Thus, for $D > 4$ the universal entanglement process with $p_4 = 0$ produces output states
with the smallest possible von Neumann entropy, namely 
\begin{equation}
S_{min} \equiv S(p_4 = 0) = {\rm ln}(D-1).
\label{D>4}
\end{equation}
For $D < 4$ this process of minimal von Neumann entropy is characterized by $p_4 = 1$ and the
corresponding minimal entropy is given by
\begin{equation}
S_{min} \equiv S(p_4 = 1) = {\rm ln}\frac{(D-1)(D-2)}{2}.
\label{D<4}
\end{equation}
For $D=3$ the resulting pure anti-symmetric output state is
orthogonal to the pure input state. Geometrically this anti-symmetric
output state may be viewed as representing the
unique plane which is orthogonal to the input state.
This way this process preserves
information about the input state. 
For $D=4$ both processes, i.e. $p_4=0$ and $p_4=1$, yield the same von Neumann entropy
for the output states. As apparent from Fig.\ref{Fig3} this possibility of a
`coexistence' of two universal optimal
entanglement processes for $D=4$ resembles some of the
signatures of a second order phase transition.
Within the one-parameter family of optimal universal entanglement processes
the process characterized
by $p_4 = (D-2)/D$ (or equivalently $C = -1/[D(D-1)]$)
gives rise to output states with the largest possible value of the
von Neumann entropy, namely 
\begin{equation}
S_{max} \equiv S(p_4 = \frac{(D-2)}{D}) = {\rm ln}\frac{D(D-1)}{2}.
\label{max}
\end{equation}
Thus this process generates an output state which is a maximal mixture
of all possible $(D-1)(D-2)/2$ anti-symmetric two-particle states.

Let us finally, determine the index of correlation 
\cite{Barnett}
\begin{eqnarray}
IC(\rho) &=& S(R_1({\bf m})) + S(R_2({\bf m})) - 
S(\rho_{out}({\bf m}))
\end{eqnarray}
of the  one-parameter family of optimal entanglement processes.
Thereby 
\begin{eqnarray}
R_{1}({\bf m}) &\equiv& {\rm Tr}_{2} \{\rho_{out}({\bf m})\} 
,\nonumber\\
R_{2}({\bf m}) &\equiv& {\rm Tr}_{1} \{\rho_{out}({\bf m})\} 
\end{eqnarray}
denote the reduced density operators of the first and second quantum
system.
This index of correlation or mutual entropy serves
as a measure for the classical and quantum correlations
between both quantum systems. 
For a given value of $p_4$ the corresponding index of correlation
is given by
\begin{eqnarray}
IC(p_4) &=& {\rm ln}\frac{4}{1+p_4} + p_4 {\rm ln}\frac{2p_4 (D-1)}{(1+p_4)(D-2)}.
\end{eqnarray}
From this relation it is apparent that $IC(p_4)$ has a local minimum
for $p_4=(D-2)/D$. Thus, the optimal entanglement process with the largest
possible von Neumann entropy produces output states with the smallest possible
mutual entropy. Furthermore, the output states of the optimal entanglement 
process with $p_4 =0$ have the largest possible index of correlation, i.e.
$IC(p_4=0) = 2{\rm ln}2$. It is remarkable that
this latter index of correlation is indendent of the dimension of the Hilbert spaces
$D$ and that this value is equal to the mutual entropy of a Bell state.

It is also of interest to investigate to which extent the optimally entangled
output states preserve information about the initial pure input state $\rho_{in}({\bf m})$.
This information about the input state
is characterized by the generalized Bloch vector ${\bf m}$.
In the output state of Eq.(\ref{ouput}) this information is contained in the terms proportional
to the parameters
$\alpha^{(1)}$, $\alpha^{(2)}$  and $\beta$. The parameters 
$\alpha^{(1)}$ and  $\alpha^{(2)}$ characterize the information about the 
initial pure input state which is still contained in the two-particle output
state in each subsystem separately, i.e. in the reduced states
\begin{eqnarray}
R_{1}({\bf m}) &=& \frac{{\bf 1}}{D} + D\alpha^{(1)}m_{ij}{\bf A}_{ij}
,\nonumber\\
R_{2}({\bf m}) &=& \frac{{\bf 1}}{D} + D\alpha^{(2)}m_{ij}{\bf A}_{ij}
\end{eqnarray}
of the first and second quantum system.
The parameter $\beta$ characterizes the information
about the input state which is distributed over both quantum systems. This latter
property is apparent from
the fact that this parameter appears  in Eq.(\ref{ouput}) with tensor products of the form
${\bf A}_{ij}\otimes {\bf A}_{jl} $ and 
${\bf A}_{ji}\otimes {\bf A}_{lj} $.
According to Eqs.(\ref{invert})
for a given value of $p_4$
these characteristic quantities are given by
\begin{eqnarray}
\alpha^{(1)}+ \alpha^{(2)}
 &=& \frac{(D-2)}{2D^2(D-1)} - \frac{p_4}{2D(D-1)},\nonumber\\
\beta+
\beta^* &=& -\frac{1}{D(D-1)} + \frac{p_4}{(D - 2)(D-1)}.
\label{optimalparameters}
\end{eqnarray}
Thus,
the universal entanglement process with $p_4 = 0$
yields
\begin{equation}
\alpha^{(1)}_{max} = (D-2)/[4D^2(D-1)]
\label{amax}
\end{equation}
and
preserves
the maximum amount of information about the
initial state in each subsystem separately.
It is instructive to compare this 
maximum value for $\alpha^{(1)}_{max}$ 
with the corresponding value achievable by
an optimal quantum cloning process which
maximizes $\alpha^{(1)}$
with respect to all possible universal quantum
processes.
Its optimal value
is given by
$\alpha^{(1)}_{clone} = (D-2)/[4D^2(D-1)] + 1/[2D^2(D-1)(D+1)]$. 
Thus, for $D>2$ both values differ by terms of relative magnitude $O(1/D^2)$ so that their
difference tends to zero rapidly
with increasing dimension  $D$ of the one-particle Hilbert spaces.
This demonstrates that for $D \gg 2$ an optimal universal entanglement process
with $p_4 = 0$  preserves almost as much
information about the orientation of the initial quantum state
as an optimal universal cloning process (compare with Fig.\ref{Fig4}).

The optimal entanglement process with $p_4 = (D-2)/D$ yields
$\alpha^{(1)}=\alpha^{(2)}= \beta = 0$
so that all information about the orientation of the initial quantum state $\rho_{in}({\bf m})$
is lost.
Its resulting output states are independent
of the input states and they are
scalars with respect to unitary transformations of the form
$U\otimes U$ and with respect to permutations between both particles.
This particular process is the only one within the
one-parameter family of optimal universal entanglement processes which fulfills the
additional requirement 
\begin{eqnarray}
R_{1}({\bf m}) = R_{2}({\bf m})  = {\bf 1}/D.
\label{Bell}
\end{eqnarray}
Though this property
is characteristic for all Bell states it does not hold for the 
output states which are generated by all other optimal universal entanglement processes with
$p_4 \neq (D-2)/D$ which are possible for $D>2$.

Let us finally comment on the special case of qubits, i.e. $D=2$, for which some of the
considerations of this chapter have to be modified. In this case $\rho_4$ disappears from
Eq.(\ref{optimal}).
Consequently only one optimal entanglement process is possible which is characterized
by $p_1 = p_3  =  0$ and by the anti-symmetric output state
\begin{eqnarray}
\rho_{out}^{(opt)} ({\bf m} \equiv D{\bf e}_{11}) = \rho_2 ^{(opt)}.
\end{eqnarray}
Thus, in this case the one-parameter family of optimal universal entanglement processes
which is possible for $D > 2$ collapses to a single process whose output states are
independent of the input states.

%

\section{Conclusions}

It has been demonstrated that in Hilbert spaces of dimensions larger than
two the linear character of quantum mechanics is compatible with the
existence of
optimal universal two-particle
entanglement processes which preserve information
about initial input states. 
These optimal universal
entanglement processes form a one-parameter family and
their resulting output states are statistical mixtures
of anti-symmetric states.
This situation is completely different from the case of qubits
where only one optimal universal two-particle entanglement process is possible
which destroys all information about any input state.
In higher dimensional Hilbert spaces
there is only one particular member of this one-parameter
family for which all information about an initial input state is lost
completely during the entanglement process. All other processes within this
family preserve this information at least partly. 
The degree of preservation of this
information about the input state
approaches the degree which is achievable by optimal universal
cloning processes. The dimensional dependence of 
the optimal universal entanglement process
whose output states have minimal von Neumann entropy exhibits signatures
of a second order `phase transition' for Hilbert space dimensions equal to four.
For this particular dimension two different
optimal universal entanglement processes
are possible which produce output states with minimal von Neumann entropy.

The presented investigations indicate that convex sums of
anti-symmetric quantum states
might also play a predominant
role in universal entanglement processes which involve more than two quantum
systems. Furthermore,
entanglement processes which also preserve 
information about an input state might have interesting applications 
in various branches of 
quantum information processing, such as quantum cryptography 
or quantum
error correction.
Thus, the presented results indicate that further exploration of
quantum information processing
beyond qubits may offer unexpected and useful surprises.

This work is supported by the
Deutsche Forschungsgemeinschaft within the SPP `Quantum Information Processing',
by the DAAD and by the DLR.

\newpage
\setlength{\unitlength}{1.0cm}
\begin{picture}(8,4.5)
\put(0.3,4){{${\bf m}_0$}}
\put(1.5,4.10){\vector(1,0){2}}
\put(4.1,4){{${\bf m} = {\bf U}{\bf m}_0$}}
\put(0.7,3.5){\vector(0,-1){2}}
\put(1.2,2.5){{${\cal P}$}}
\put(5,2.5){{${\cal P}$}}
\put(4.6,3.5){\vector(0,-1){2}}
\put(0,1){{$\rho_{out}({\bf m}_0)$}}
\put(1.5,1.10){\vector(1,0){2}}
\put(4.2,1){{$\rho_{out}({\bf m}) =$}}
\put(4.2,0.5){{${\bf U}\otimes {\bf U}
\rho_{out}({\bf m}_0){\bf U}^{\dagger}\otimes {\bf U}^{\dagger}$}}
\end{picture}
\begin{figure}
\caption{Pictorial representation of the symmetry (covariance) condition
which characterizes universal quantum processes}
\label{Fig1}
\end{figure}

\centerline{\psfig{figure=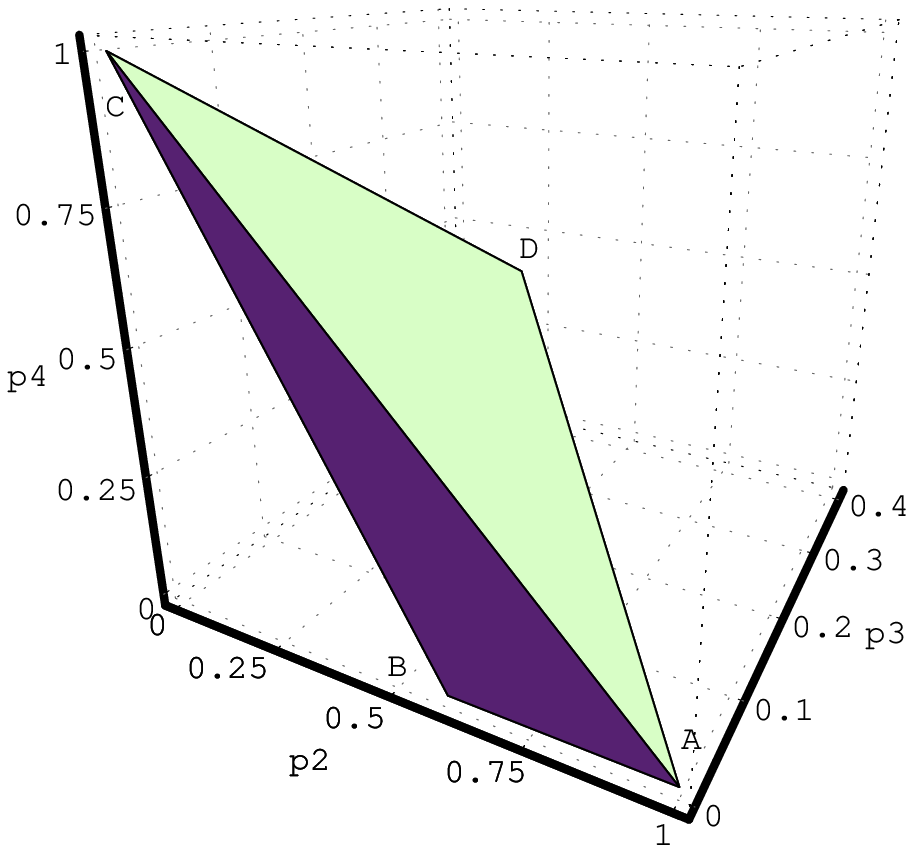,width=8.6cm,clip=}}
\begin{figure}
\caption{
Convex set of points $(p_2,p_3,p_4)$
characterizing all possible universal quantum processes for 
$\alpha^{(1)}=\alpha^{(2)}$,
$\beta = \beta^*$ and 
$D=4$.}
\label{Fig2}
\end{figure}

\centerline{\psfig{figure=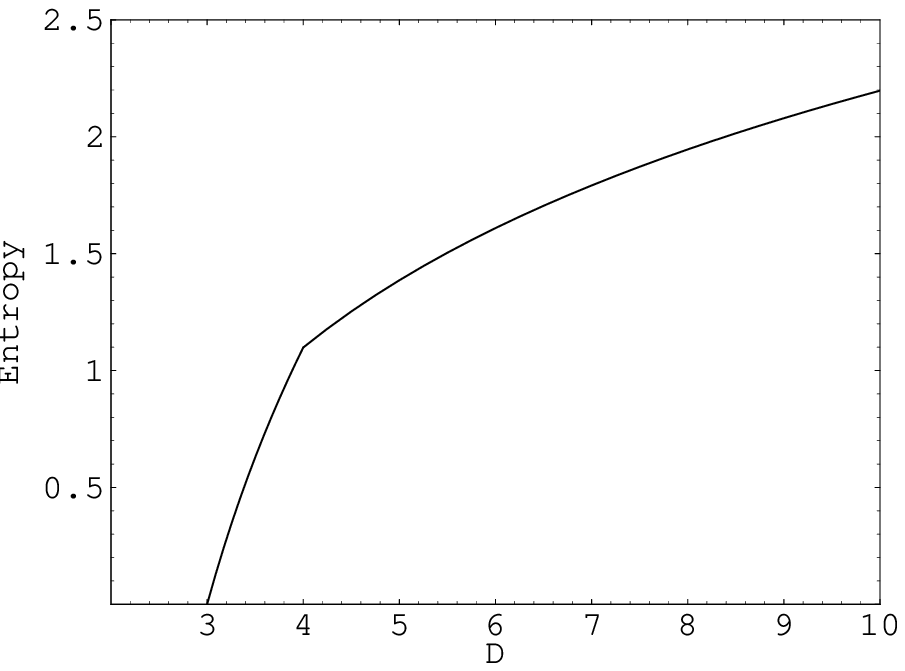,width=8.6cm,clip=}}
\begin{figure}
\caption{Minimal values of the von Neumann entropy of optimal universal entanglement processes
(compare with Eqs.(\ref{D>4}) and (\ref{D<4}))
as a function of $D$.}
\label{Fig3}
\end{figure}

\centerline{\psfig{figure=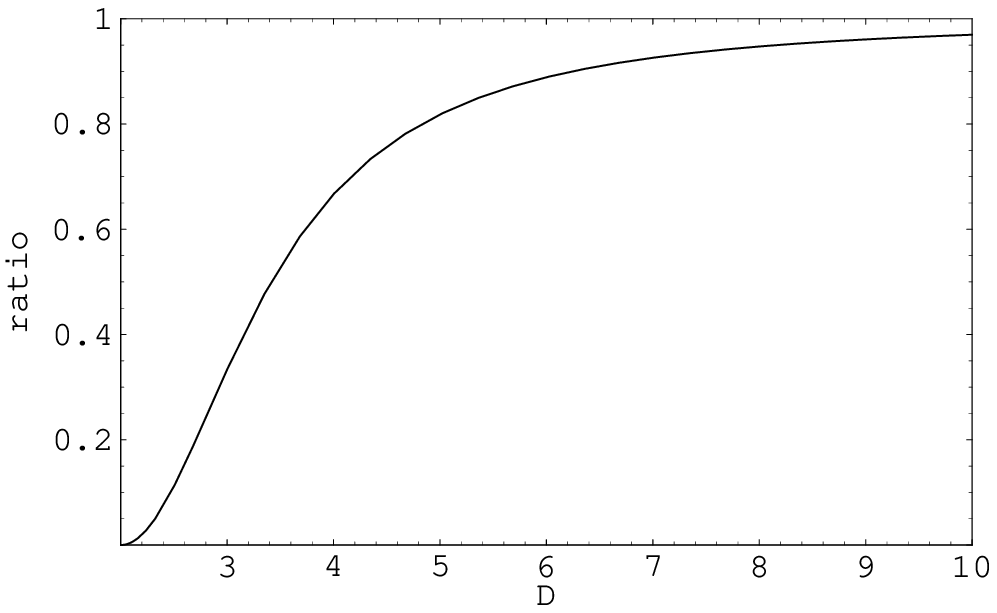,width=8.6cm,clip=}}
\begin{figure}
\caption{ Dimensional dependence of the ratio between
$\alpha^{(1)}_{max}$ as defined by Eq.(\ref{amax}) and the corresponding value
$\alpha^{(1)}_{clone}$
characterizing the optimal universal cloning process.
It is for $D=2$ only that in the optimal universal entanglement process all
information about any input state is lost.}
\label{Fig4}
\end{figure}

\end{document}